\documentclass[runningheads]{svmult}

\usepackage{makeidx}   % allows index generation
\usepackage{graphicx}  % standard LaTeX graphics tool
\usepackage{subeqnar}  % subnumbers individual equations
\usepackage{multicol}  % used for the two-column index
\usepackage{physprbb}  % modified textarea for proceedings,
\makeindex             % used for the subject index
\newcommand\ha{H${\alpha}$~} %letizia
\newcommand\sm{M$_{\odot}$} %letizia

\begin{document}
\title*{Clues on post-Asymptotic Giant Branch Evolution and Planetary Nebulae
Populations from the Magellanic Clouds.}
%%
% allows explicit linebreak for the table of content
%
%
\titlerunning{Magellanic Cloud PNe}
% allows abbreviation of title, if the full title is too long
% to fit in the running head
%
\author{Letizia Stanghellini}
\authorrunning{Letizia Stanghellini}
% if there are more than two authors,
% please abbreviate author list for running head
%
%
\institute{National Optical Astronomy Observatory
950 N. Cherry Ave., Tucson AZ 85719, USA}

\maketitle              % typesets the title of the contribution

\begin{abstract}
%The abstract\index{abstract} is optional. If present it should summarize
%the contents of the paper in at least 70 and at most 150 words; neither
%too long nor too short but to the point!

\end{abstract}
The recent {\it HST} 
optical images, and the optical and ultraviolet spectra, of Magellanic planetary nebulae (PNe),
together with the large data-base that has been collected in the past decade, 
allows unprecedented insight in the evolution of PNe and their central stars. 
In this paper
we present a selection of recent results: The analysis of PN
morphology, both in the optical and ultra-violet emission lines; the relation between 
nebular morphology and the 
chemistry produced by stellar evolution; 
the direct determination of the transition time from observations;
and the study of the nature and evolutionary stage of the components of the planetary nebula luminosity function.

\section{Introduction}
Planetary nebulae (PNe) in the Magellanic Clouds (LMC, SMC) are the nearest
extragalactic PNe known. They are close enough for detailed individual analysis, and they 
can be spatially resolved with the current technology; yet, they do not suffer the distance biases and extreme differential 
interstellar absorption that Galactic PNe are well known for. Magellanic 
PNe are thus the ideal probes to study stellar evolution of low- and 
intermediate-mass stars. They also are ideal benchmarks to extend the knowledge
of PNe to the extra-Galactic PN populations, with the additional bonus of
covering a large metallicity range.

PNe in the Magellanic Clouds have been observed from the ground, as point sources, since the
sixties. The advent of the Hubble Space Telescope ({\it
HST}) afforded their spatial resolution, making it possible 
to study their 
size, morphology, and their central stars (CSs).
The work presented in this paper is based on the most recent {\it HST} data-sets acquired by the MCPN team\footnote{
Team membership and other information are available in the MAST MCPN web page http://archive.stsci.edu/hst/mcpn/home/html} in Cycles 8 through 10 with
the Space Telescope Imaging Specrograph (STIS, Programs 8271, 8663, 9077, and 9120) and the Wide Field Planetary Camera 2 (WFPC2, Program 8702). 
Older archived narrow-band images (Program 6407, 5185, 4821, 4075, and 1266) were also used to augment the morphological
data-set.

\begin{figure}
\centering
 \includegraphics[width=3.8cm]{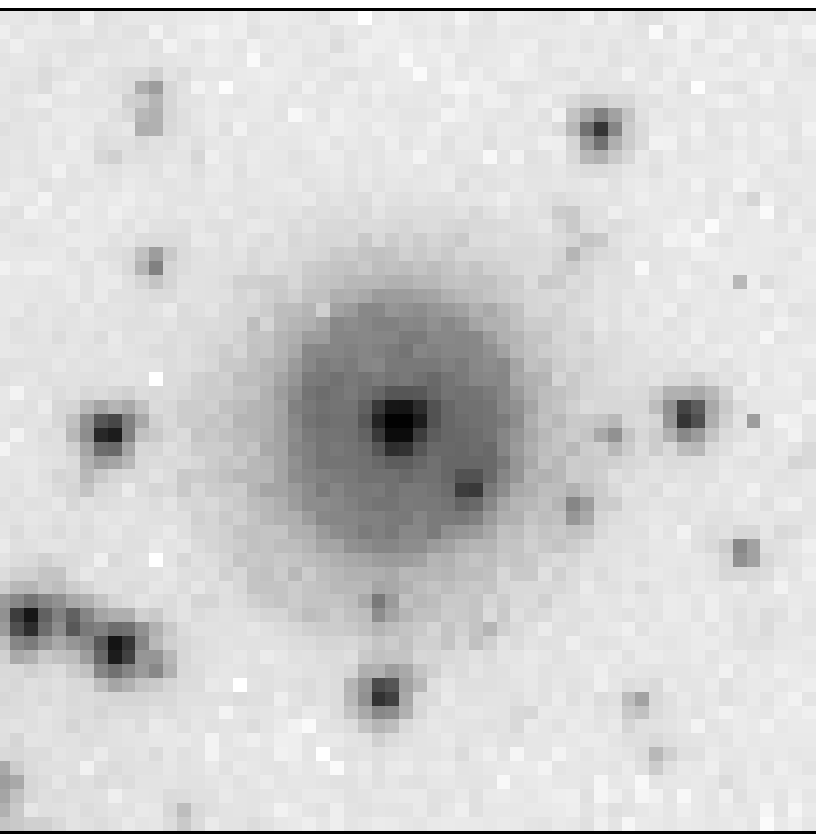}%
% \hspace{1cm}%
 \includegraphics[width=3.8cm]{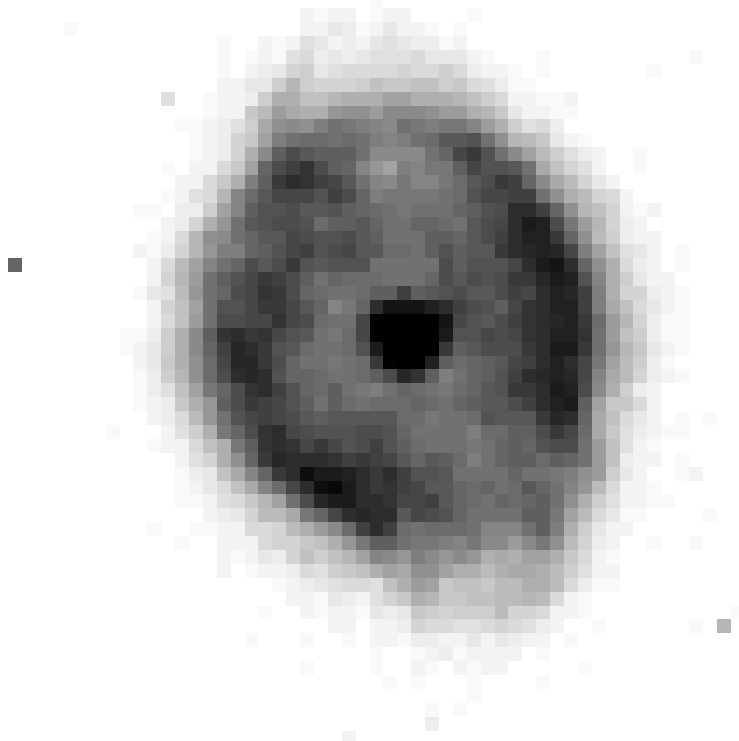}%
%\hspace{1cm}%
 \includegraphics[width=3.8cm]{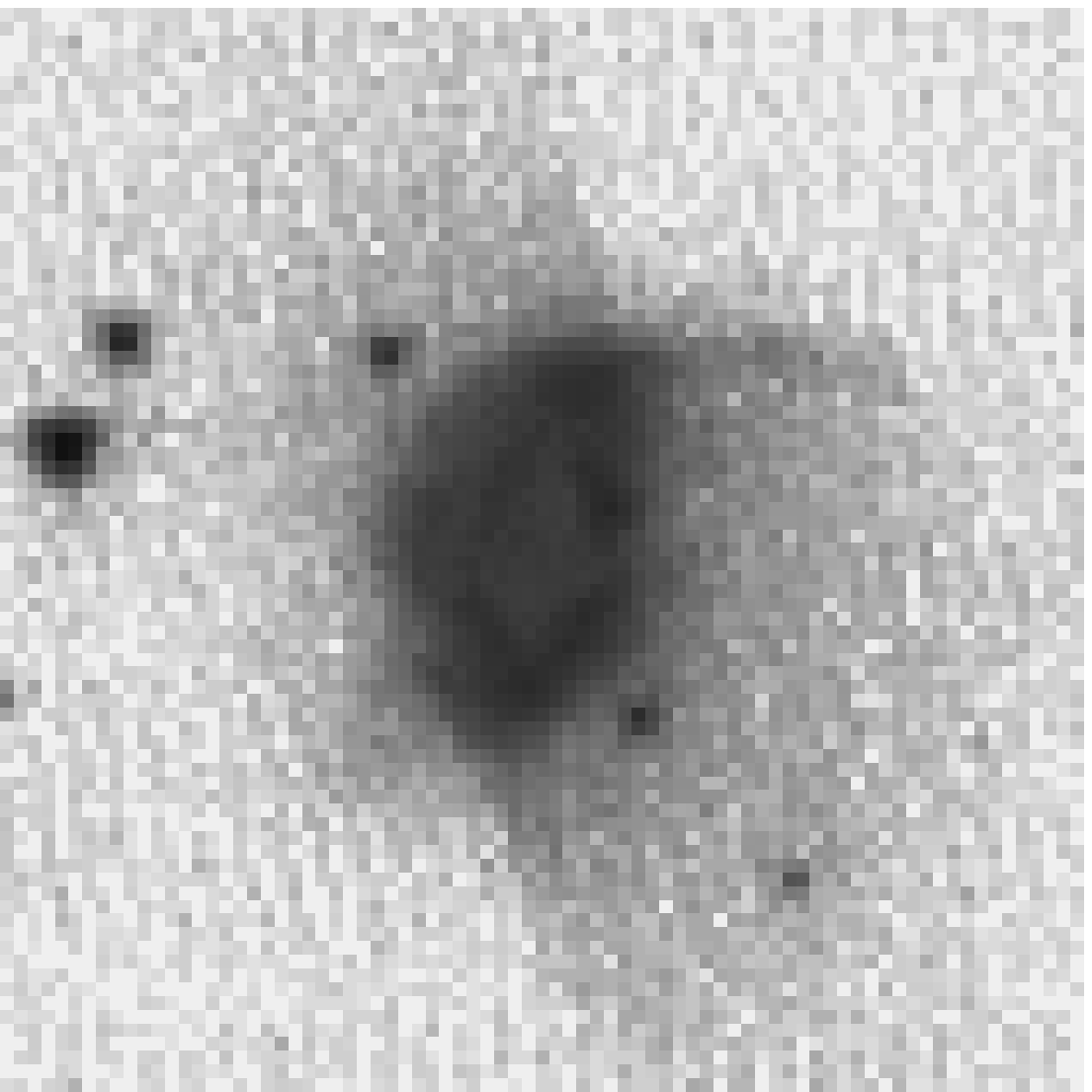}%
\hspace{1cm}%
 \includegraphics[width=3.8cm]{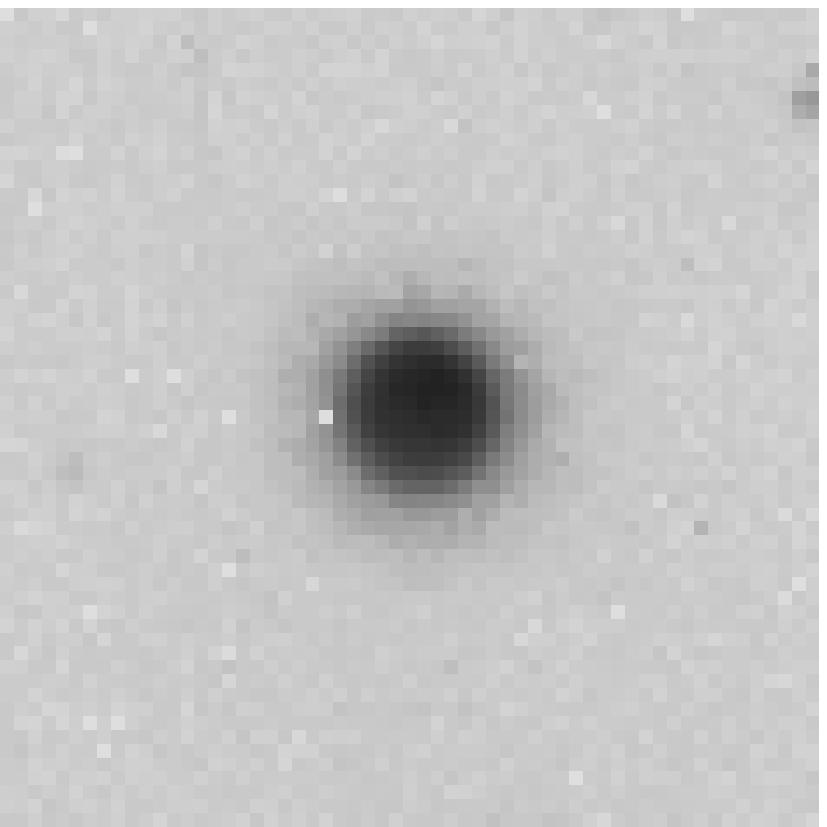}%
%\hspace{1cm}%
 \includegraphics[width=3.8cm]{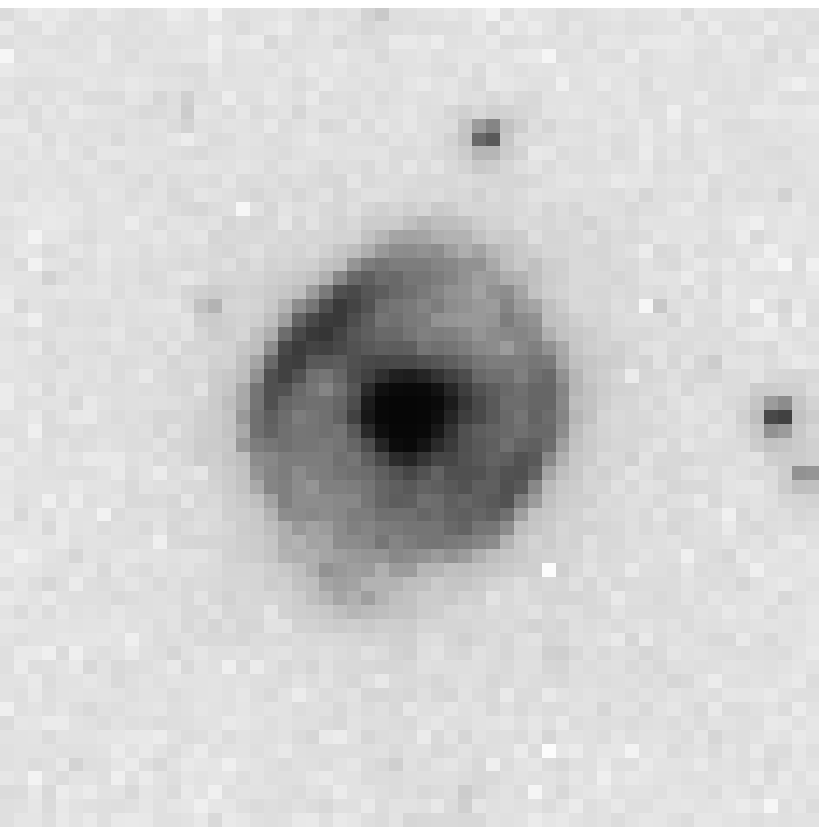}%
% \hspace{1cm}%
 \includegraphics[width=3.8cm]{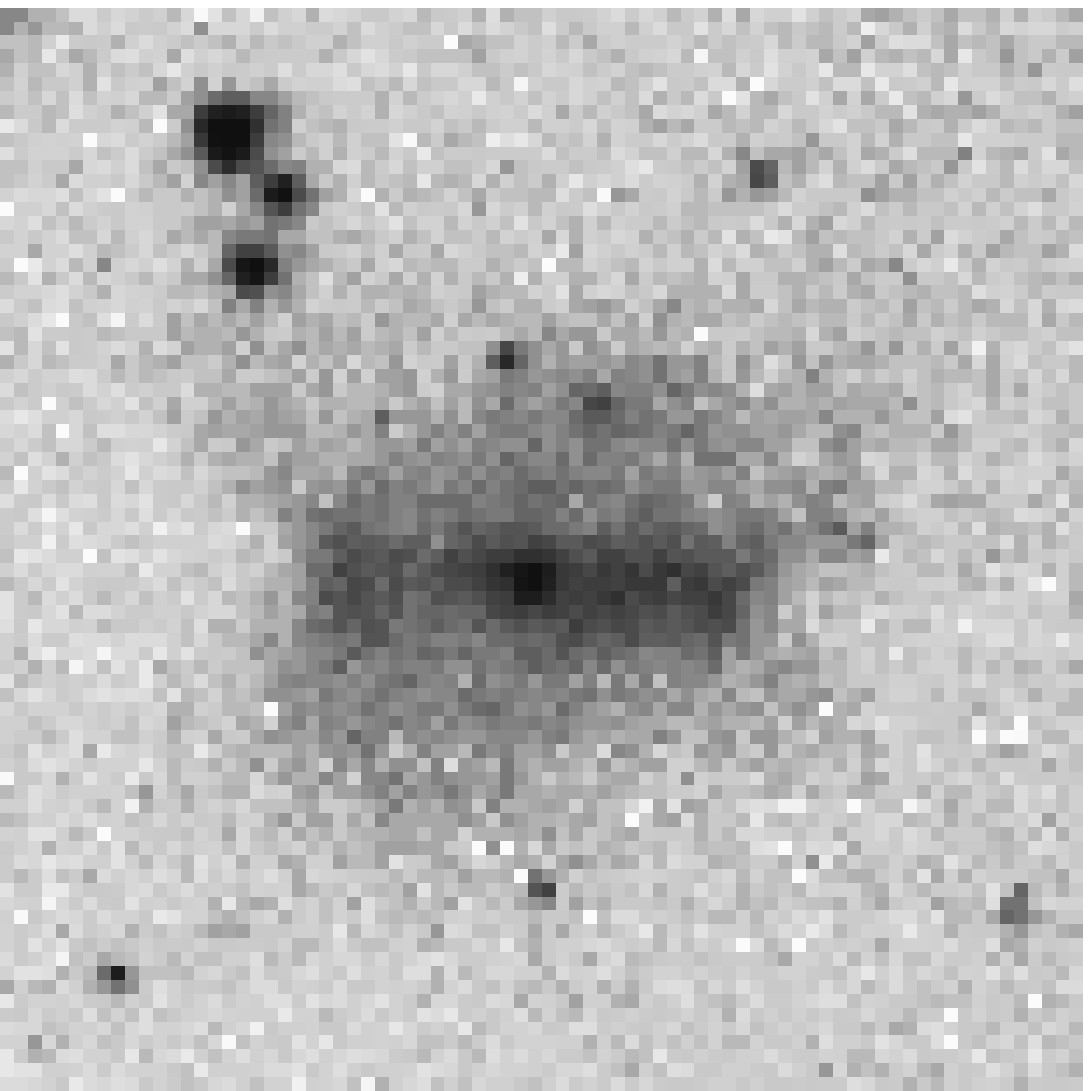}%
  \hspace{1cm}%
 \includegraphics[width=3.8cm]{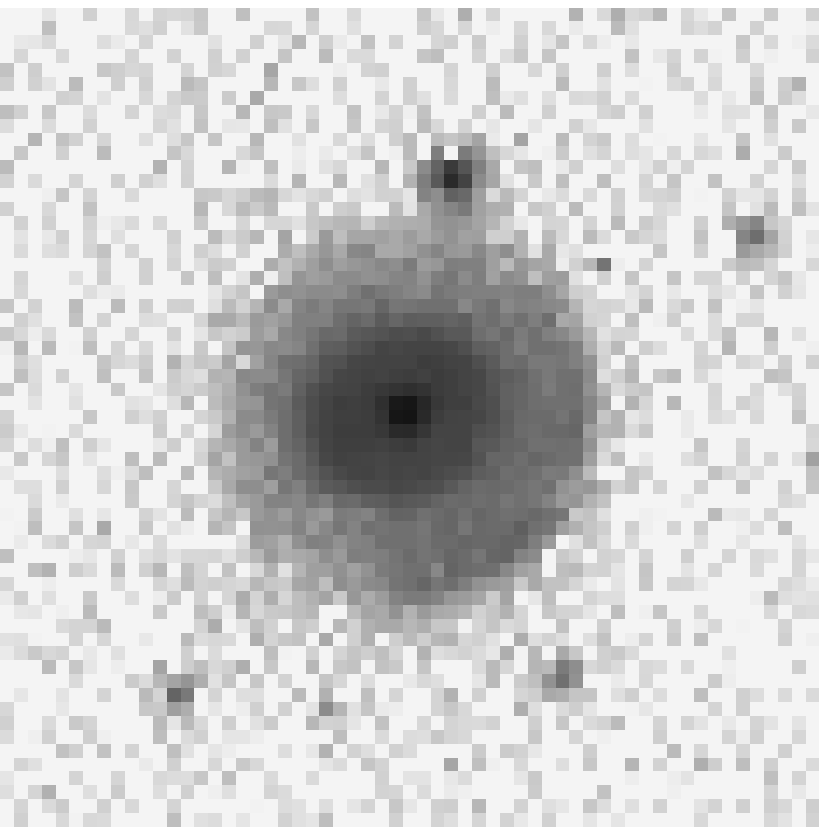}%
% \hspace{1cm}%
 \includegraphics[width=3.8cm]{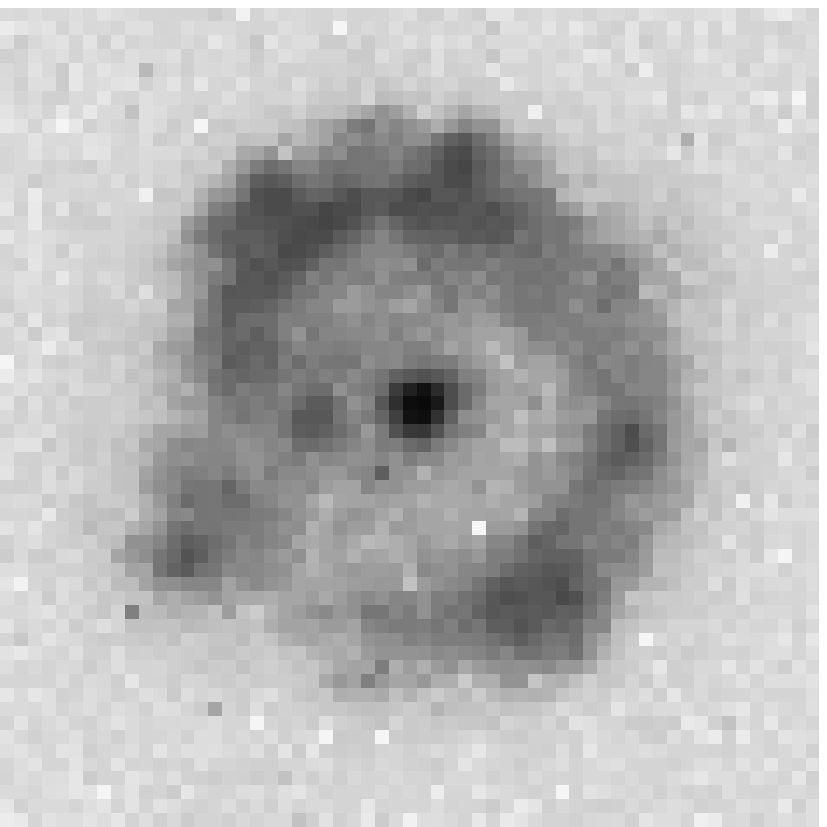}% 
% \hspace{1cm}%
 \includegraphics[width=3.8cm]{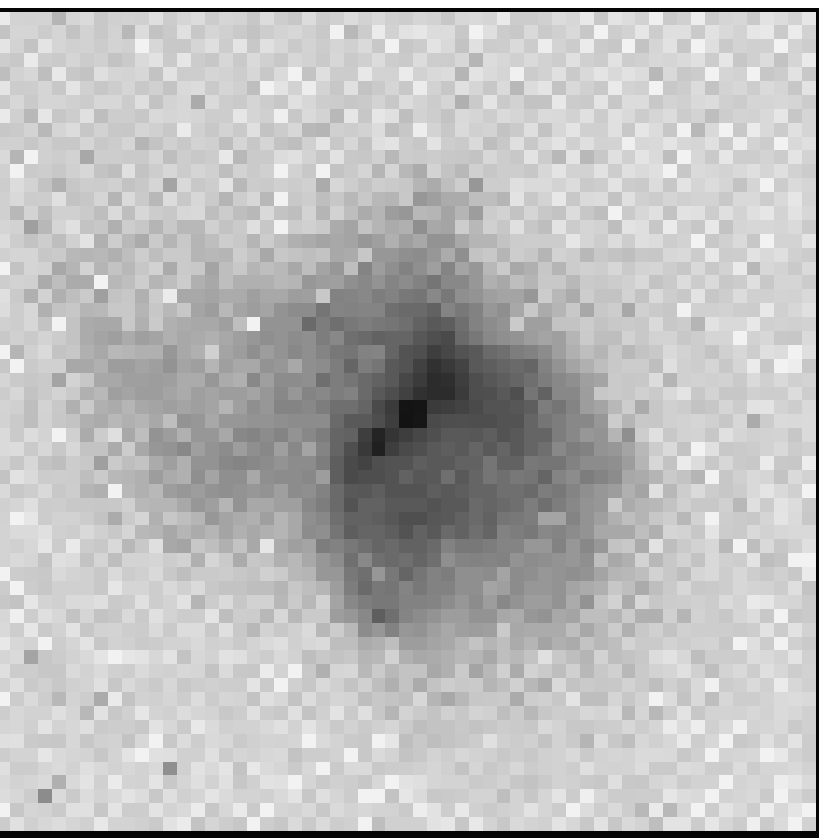}% 

    \caption{Magellanic PNs. Left panel (each panel from top to bottom): R PNs LMC~J~33, 
    SMC~SP~34, and LMC~MG~40; Central panel: E PNs LMC~SMP~101, SMC~MG~8,
    and SMC~MG~13; Right panel: B PNs  LMC~SMP~91, SMC~MA~1682,
    and LMC~MG~16. All thumbnails are 9 arcsec$^2$ sections of the
    STIS broad-band images
    }
\end{figure}

\section{Planetary Nebula Morphology}

The morphology of Galactic PNe has been studied thoroughly in the nineties,
and it has been found that morphological type is related to
the PN progenitor's evolutionary history and its mass. There is strong evidence
that bipolar PNe are the progeny of the massive
Asymptotic Giant Branch (AGB) stars (4-8 \sm). Bipolar PNe are nitrogen-rich and carbon-poor
[1,2]. The analysis of the morphological 
types and their distribution in a given PN population are then good indicators of age and evolutionary history.

Galactic PNe have been classified as round (R), elliptical (E), bipolar (B; includes quadrupolar and multipolar), bipolar
core (BC; R or E PNe with a central {\it bilobate} concentration, or ring enhancement), 
and point-symmetric (P). A recent description of these classes is published by [1].
The majority of Galactic PNe are elliptical, but the actual number of
B PNe could be underestimated,
given that they typically lie in the Galactic plane (i.e., they may suffer from high 
reddening).
PNe in the Magellanic Clouds, when spatially resolved, show the same admixture of morphological types 
than the Galactic PNe, as illustrated in Figure 1 [3, 4, 5].
While we do not attempt a statistical comparison of the
Magellanic and Galactic PN morphological types, given the selection effects that hamper
Galactic PNe, we can meaningfully compare the LMC and SMC samples. Both
samples are characrerized by low field extinction, and they have been preselected in
more or less the same way.

\begin{table}

\caption{Morphological Distribution}
\begin{center}
\begin{tabular}{lrr}
&&\\
Morphological type& $\%$ LMC&  $\%$ SMC\\
&&\\  
Round (R)&		29&	35\\
Elliptical (E)&		17&	29\\
R \& E&	                46&	       64\\
Bipolar (B)&		34&	6\\
Bipolar core (BC)&	17&	24\\
Point-symmetric (P)&	 3&	6\\	

    \end{tabular}

\end{center}
\end{table}

The results of the morphological distribution of PNe in the Clouds is illustrated in Table 1.
One striking difference 
between the LMC and SMC morphological type distributions is that the fraction of
B PNe in the LMC is almost six times that of the SMC. Furthermore, half of the LMC
PNe are either bipolar or have a morphology that suggests a central ring (BC),
while only a third of the SMC PNe fall in these two aspherical morphological classes.

From this analysis it results that the different processes involved in the 
formation of the
different PN shapes occur in all galaxies where morphology has been
studied. It is also apparent that the SMC environment may disfavor the onset
of bipolarity in PNe. Otherwise, the different morphological statistics 
may indicate different populations of stellar progenitors in the two Clouds.
While it seems reasonable to conclude that a low metallicity environment 
is unfavorable to bipolar formation and evolution, the exact causes have not been studied
yet. A detailed study of metallicity and mass-loss may clarify this 
point. On the other hand, the different morphology counts may simply 
imply that most of the progenitors of the SMC PNe have masses in the lower-mass range (M$<$4\sm),
thus they do not produce B PNe.
A comparison of the SMC and LMC CS masses does not disagree with such a possibility [6].

\begin{figure}
\begin{center}
 \includegraphics[width=.8\textwidth]{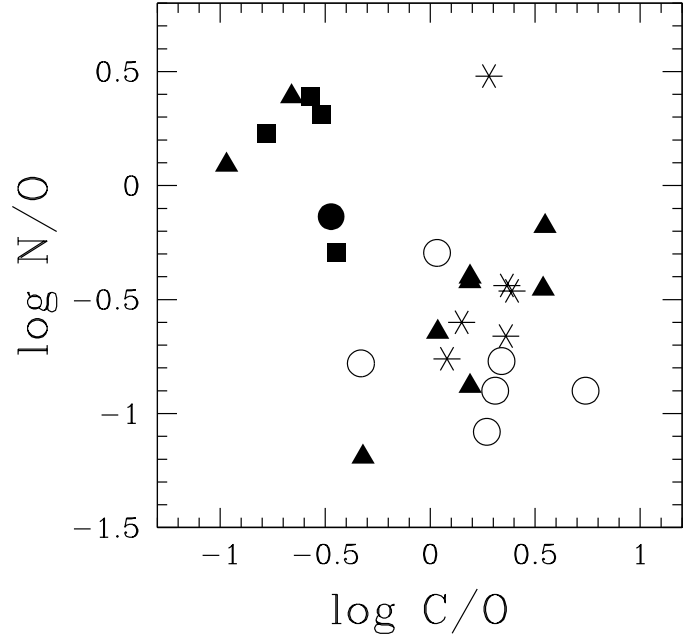}
 \end{center}   
 \caption{The C/O vs. N/O relation for R (open circles), E (asterisks),
BC (triangles), B (squares), and P (filled circles) PNe in the LMC
(adapted from [11])}
\end{figure}

\section{Results from UV Spectroscopy}

The study of the chemical composition that characterizes PN ejecta 
is a way to determine the origin and evolution of the progenitor stars. 
PNe contain the products of stellar nucleosynthesis, dredged-up at the
stellar envelope during the late AGB evolution. The study of carbon and nitrogen 
abundance in PNe is especially interesting. The theory of stellar evolution 
suggests that Galactic and Magellanic AGB evolution 
yield to different N and C concentrations depending on the AGB progenitor's mass. 
If M$_{\rm MS} >4$ \sm, the
so-called {\it hot-bottom burning} (HBB) process would convert much of the 
carbon content into nitrogen, resulting in a different chemical
make-up of the remnant PN [7]. In order to disclose possible correlation between the 
morphology and the {\it evolutionary} chemistry of the LMC PNe 
we need to have at our disposal the carbon and nitrogen abundance of a sufficient number of  
PNe to populate the major morphological classes. We use the nitrogen and oxygen abundances in the
literature [8, 10] and the carbon abundance in [9, 11] to plot, 
in Figure 2, the relation between the (logarithmic) C/O and N/O abundances.

From this Figure we infer that the carbon stars progeny (log C/O$>$0) is limited to R,
E, and BC PNe. B and P PNe originate from the
evolution of stars that did not go though the carbon star phase, or, more likely, their 
progenitor underwent a carbon depletion, possibly due to the occurrence of the HBB. It appears that
R and E PNe are the progeny of lower-mass stars, those with M$_{\rm MS}< 4$ \sm,
while the opposite holds for the most aspherical (B and P) PNe.
From Figure 2 we see that BC PNe might have evolved from carbon stars in some cases, while in other
cases they had evolved from more massive stars. It remains
to be seen whether some of the BC PNe are indeed bipolar with faint lobes 
(those with low carbon and high nitrogen content), and others are misclassified R and E PNe.

\begin{figure}
\centering
 \includegraphics[width=\textwidth]{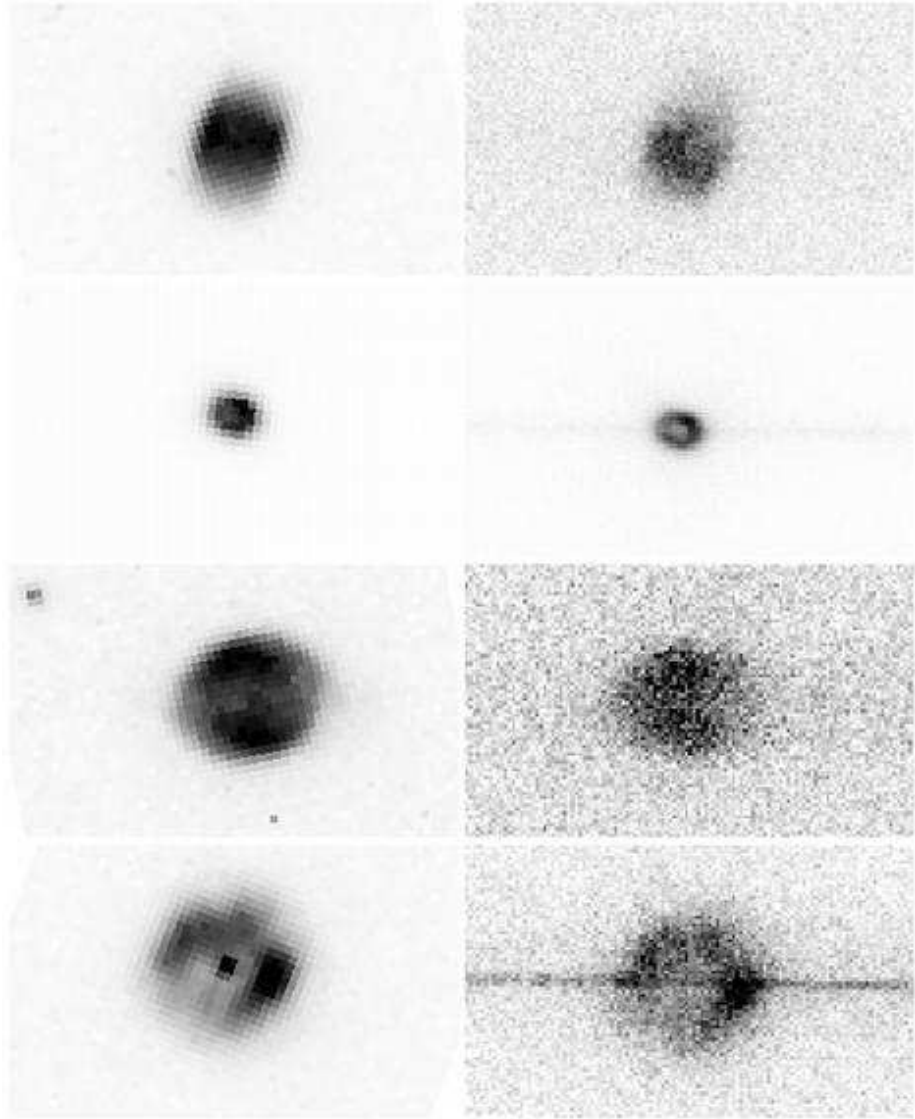}
    \caption{Morphology of LMC PNe. Lerft column: clear STIS CCD images.
    Right column: C III] 1908 \AA~images from the G230L 2D STIS spectra. The PNe are, from top to bottom, SMP~19, SMP~79, SMP~95, 
and SMP~102 }
\end{figure}

In order to enlarge the sample of Magellanic PNe whose carbon abundance is known, 
we have used the {\it HST} data
from program 9120, acquired with {\it STIS} grisms G140L and G230L [11]. The 2D spectra were obtained
with a large aperture, showing the morphology of the PNe through the brightest UV 
emission line. It is then possible to compare the UV to optical morphology. 
In Figure 3 we show the clear optical images through the {\it STIS} CCD
(left column) and the parts of the G230L spectra corresponding to the C III] emission line 
at $\lambda$1908 \AA. The image sections are all to scale, thus the (relative) physical
sizes are also to scale. The clear images have been rotated so their orientation coicides with the orientation of
the 2D spectra. The UV emission deriving from the CIII] line originates
in the same volume of the nebulae than the optical emission
[4]. In Figure 3 we can also clearly see the spectrum of
the central star of SMP~102.

The fitting of the UV stellar continua have been used to derive stellar tempeatures. In some cases,
central stars of Magellanic PNe show P-Cygni profiles in the UV spectra [12].

\section{Stellar Evolution Beyond the AGB, and the Transition Time}

The known distance of Magellanic PNe makes them the ideal probes to study stellar evolution. 
AGB and post-AGB evolution have been studied theoretically and observationally 
in many different fashions. Nonetheless, there are still several processes
that need to be understood to make the comparison between theory and data meaningful. 
In this section we describe how Magellanic PNe are used to shed some light on the
transition time problem. Other aspect of AGB evolution
will be examined in future papers.

\begin{figure}
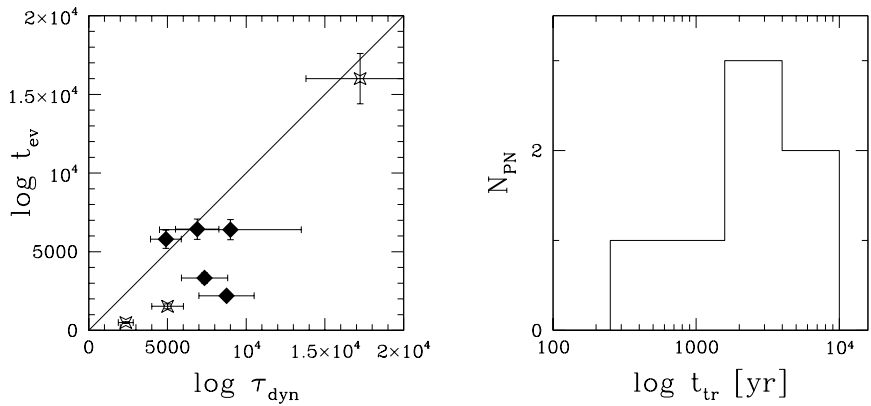

%\centering
 \includegraphics[width=.5\textwidth]{fig4a.epsi}
 \includegraphics[width=.5\textwidth]{fig4b.epsi}

 \caption{{\bf (a)}: Evolutionary vs. dynamical time for LMC (diamonds) and SMC (stars) PNe. {\bf (b)} : 
 Transition time histogram}
\end{figure}

When comparing CS data to theoretical predictions, it is often assumed that the 
time lag between the envelope ejection quenching (EEQ) and the PN illumination, the so-called transition time 
(t$_{\rm tr}$), has a duration either 
of a fixed value (generally 1000 yr), or it is set to zero. Both of these assumptions are inadequate to 
express a time-scale that is likely to assume different values in different PNe, and that represents a sizeable fraction of
the PN life-time [13]. From a conceptual point of view, the transition time 
is mainly dependent on the residual envelope mass that survives the envelope ejection (M$_{\rm e}^{\rm R}$) and 
on the post-AGB mass-consumption process (or processes) that is in place in each specific CS.
The dynamical time, $\tau_{\rm dyn}=D_{\rm PN}/v_{\rm exp}$ (where D$_{\rm PN}$ is the diameter of the nebula and 
v$_{\rm exp}$ is the nebular expansion velocity), represents the dynamical nebular age in the assumption that
the nebular shell does not suffer acceleration. If we measure the diameter of the main nebular shell, 
the dynamical time tracks the evolutionary time measured from the EEQ event. In order to compare the dynamical 
time to the stellar
evolutionary time we need to use the equation $\tau_{\rm dyn}=t_{\rm tr}+t_{\rm ev}$, where t$_{\rm ev}$ is the
evolutionary time evaluated {\it after} the PN illumination. This equation holds as long as the evolutionary 
tracks mark the temporal zero-point at
 T$_{\rm eff}\simeq 30,000$ K, as they usually do.

In order to estimate the transition time one needs to determine the residual envelope mass, the mass-loss 
during transition, and whether the CSs are burning hydrogen or helium. All these properties are rarely known, if ever.
Another approach to determine, or at least constrain, the transition time is to use the 
Magellanic CSs. Since the distance to the Magellanic PNe is known, their CS mass can
be estimated much more reliably than for Galactic PNe. In Figure 4 {\bf (a)} we plot the evolutionary time
of the Magellanic CSs, derived from their location in the HR diagram,
against the dynamical time of their nebular shells. 
Stellar luminosities and temperatures used to deterine the CSs location on the HR diagram with
respect to the Magellanic post-AGB evolutionary tracks are from [6] and [14]; the nebular diameters
are from [4] and [5], and the 
expansion velocities are taken from [15]. Note that the majority of LMC post-AGB evolutionary tracks 
are He-burning tracks [16], while
there is an indication that these stars are H-burners (see below in the PNLF section).
To compensate for this mismatch we have scaled the H- versus He-burning tracks and
increased the evolutionary time-scales accrodingly. Figure 4 {\bf (a)} shows that the dynamical and
evolutionary times correlate to one another (R$_{\rm xy}$=0.88), yet the dynamical time is usually higher, as predicted
in most evolutionary cases [13]. There is one exception, 
LMC~SMP~13, whose transition time is just below zero ( t$_{\rm ev}$ and t$_{\rm dyn}$ are
within their respective errorbars).
From the evolutionary and dynamical times we derive the transition time in all cases but for SMP~13, whose
transition time is obviously close to zero, or, alternatively, whose evolution occurs in
thermal time-scale [13]. In Figure 4 {\bf (b)} we report the histogram of
the transition times for the Magellanic PNe whose dynamical and evolutionary times are
available.  The values of the transition time spread from a few hundreds to several thoudands years. In a
few cases it goes to 10,000 yr, certainly a considerable fraction of the PN life-time.
The most important conclusion from this exercise is that transition time can not be 
ignored when dealing with the evolution of PNe and their central stars.
To our knowledge this is the first time that transition time has been effectively
estimated for nebulae whose distance is reliable. A detailed analysis of the Magellanic PN 
estimated transition times and their models will be presented in a future paper.

\section{The Planetary Nebula Lumonsity Function }

The planetary nebula luminosity function (PNLF) is a well-known secondary distance indicator [17]. 
The high-luminosity cut-off of the PNLF seems to be invariant for galaxies
with similar metallicites, while with a metallicity-correction it is also possible to unify the 
method for all galaxies. Despite its empirical success, it is still theoretically unclear why the PNLF
is invariant for different PN population. Moreover, the evolutionary mechanisms behind the PNLF shape 
are still uncertain. In particular, the PNLF seems extremely hard to model. A realistic representation
of the PNLF would involve several
unknown and unconstrained parameters and processes related to the post-AGB and nebular
evolution, such as the transition time, the mass-loss and its rates, the onset of PN
morphology. 
The morphological data-base of Magellanic PNe may be used to gain insight in the PNLF. In Figures 5 and 6 we plot the luminosity of the LMC and the 
SMC PNe against their photometric physical radii. 
The luminosity is given in [O III] $\lambda$5007 \AA~ magnitudes, as typical for PNLFs, and
the radii are in parsecs. All targets have been coded for morphological type, as described in the labels.

\begin{figure}
\centering
\includegraphics[width=.9\textwidth]{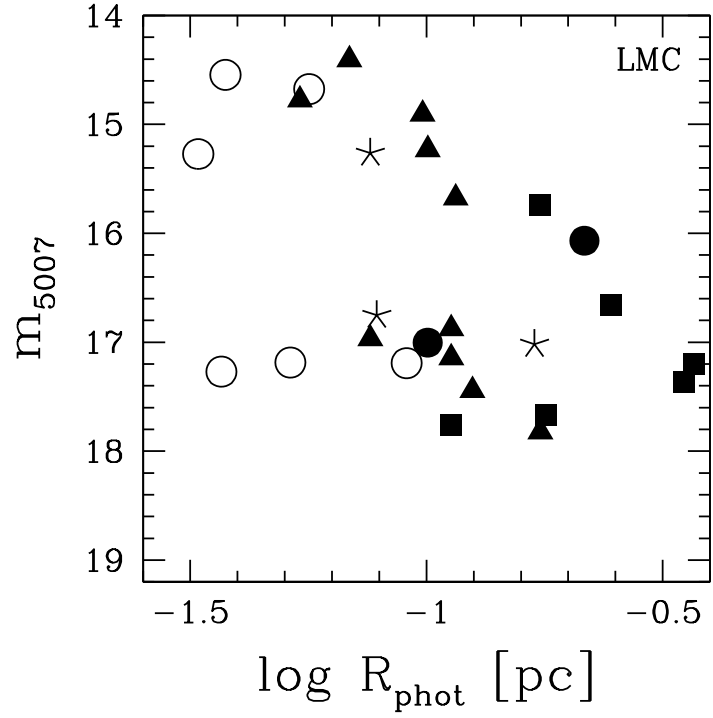}
 \caption{Magnitude $\lambda$5007 \AA~ versus photometric radius for round (open circles), elliptical (asterisks),
round and elliptical with a bipolar core (triangles), bipolar (squares), and point-symmetric
(filled circles) PNe in the LMC }
\end{figure}

The snapshot survey targets plotted in Figures 5 and 6 were selected from the brightest
LMC and SMC PNe [4, 5], thus the completeness near the 
high luminosity cut-off should be adequate [17]. 
In both Figures, but more clearly in Figure 5, one can see two different {\it branches} of PNe, 
running from small to large radii. These branches decline
in brightness for larger radii, and the B and P PNe populate the low brightness- large radii
parts of the plots.
The PNLF structure that derives from the Figure has a dip at intermediate magnitudes, as predicted
for H-burning post-AGB stars [13, 18]. 

More importantly, the high luminosity PNe in both Figures are 
R, E, and BC. These are tipically the low nitrogen
and high carbon PNe, the progeny of the lower mass stars
(M$<$4 \sm). The usefulness of this information is multi-fold: when modeling the high luminosity parts of the PNLF one
could set aside the more difficult modeling of bipolar PNe, since spherical and ellipsoidal
approximation should suffice. Furthermore, the mass range of PN progenitors that involve the
high luminosity cut-off of the PNLF may be limited to M$<$4 \sm~ in PNLF models.

The brighest PNe in the SMC are definitely R and E. The brightest B PNe are almost two magnitudes below the
high luminsoity cut-off. In the LMC, there is a considerable
fraction of bright BC PNe. Are these the same BC PNe
with high carbon and low nitrogen abundances as seen in Figure 2? We have abundance information for three of the five
bipolar core PNe in the brightest two magnitudes of the LMC PNLF, and these three objects have high carbon and low
nitrogen content. 

In Figure 6 we plot all the known (spatially-resolved) SMC PNe and
the two H II regions corresponding to young stellar clusters, that were previously misclassified as
PNe [19]. 
The Figure illustrated that misclassified PNe may occupy the same areas of the diagnostic m$_{\rm 5007}$ - log R$_{\rm phot}$
plot.
The brightest of these miscalssified PNe is just one magnitude below the high luminosity 
cut-off of the SMC PNLF. In a galaxy particularly rich of these ultra-compact H II regions that embedd young stellar
clusters the distance
determination based on the PNLF should be revisited after accounting for the contamination effects.
 
\begin{figure}
\centering
\includegraphics[width=.9\textwidth]{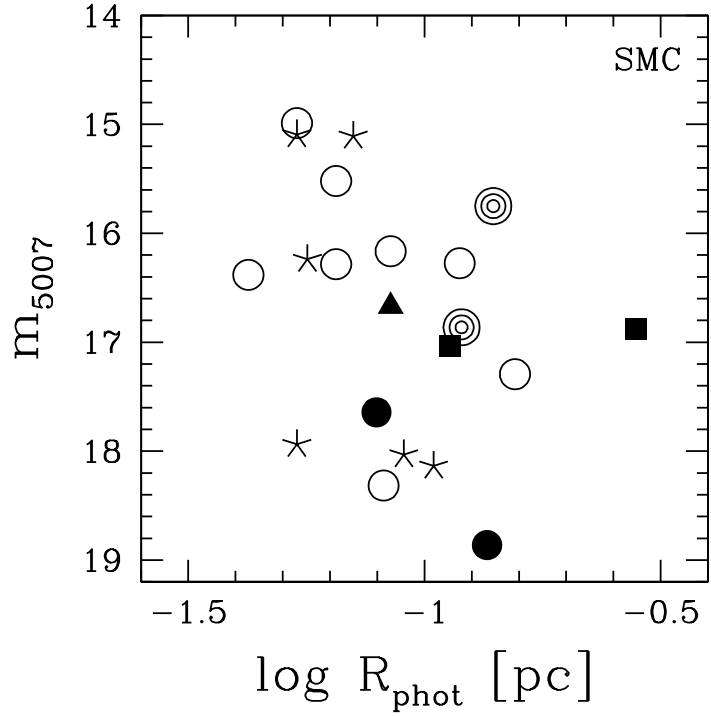}
 \caption{Same as in Figure 5, for SMC PNe. The location of the target-like symbols correspond to
 two previously misclassified ultra-compact H II regions}
\end{figure}

\section{Finale}

Magellanic PNe and their central stars have proven essential to probe post-AGB evolution and 
the ensemble properties of PNe, such as those related to the PNLF.
The multi-wavelength 
data-base of Magellanic PNe is homogeneous, and now comprises spatially-resolved {\it HST} images
of a represenaive fraction of PNe in these galaxies. This data-base 
allows umprecedented
insight of stellar and nebulae evolution. 
In this paper we showed that R, E, B and P nebular morphology
exists in both the LMC and the SMC, and that their relative frequency is different, being aspherical
PNe rare in the SMC. Morphology in the light of the ultraviolet C III] $\lambda$5007 \AA~ line is similar to
that of the major optical emission lines, such as \ha and [O III] $\lambda$5007 \AA~. 

We have also shown that, as in the Milky Way, morphology correlates
with nitrogen and carbon abundances, 
and that the aspherical PNe are typically carbon-poor and nitrogen-rich, as expected of the
progeny of the high-mass AGB stars. 

By comparing rate of nebular and stellar evolution of  Magellanic PNe it has been possible to measure 
the time-lag between the EEQ and the PN illumination. The resulting transition time are a sizeable fraction of 
the nebular life-time in some cases. We conclude that the transition time can not be ignored when comparing nebular and stellar models.

The Magellanic PNLF can be studied in its components. The high luminosity PNe are generally
R or E, and, in all the cases where abundances have been calculated,
the high luminosity PNe are carbon-rich and nitrogen-poor. This result has bearing on the
future modeling of the PNLF.

Some of the results outlined here will be more sound when the planned Magellanic PN data-set 
will reach completion. In particular, we
plan to obtain UV spectra of several SMC PNe to extend the result of PN carbon abundances to the SMC. The transition time calculation will be extended to the whole sample of Magellanic PNe
whose central stars have been observed, and more SMC central
stars will be observed as planned. 
Finally, a thorough study of the Magellanic PNLF is about to be completed by the
MCPN team with the inclusion of the optical images form recent {\it HST} observations. The characteristics of the central stars hosted by the brightest nebulae in each galaxy will be studied in detail for more 
astrophysical insight in the invariance of the PNLF high luminosity cut-off
across galaxies.

\vskip 1truecm

It is a pleasure to thank the MCPN team for their contribution to
this project. Arturo Manchado and the IAC are warmly thanked for their hospitality in July 2004, when this paper was written.

\end{document}